\documentclass[conference, 1pt]{IEEEtran}

\usepackage{algorithm}
\usepackage{algorithmic}
\usepackage{amsmath, amssymb, graphicx}

\IEEEoverridecommandlockouts 
\begin{document}
%
\title{ANM-PhaseLift: Structured Line Spectrum Estimation from Quadratic Measurements}

\author{\IEEEauthorblockN{Zhe Zhang, Zhi Tian}\thanks{This work was partly supported by the NSF grants \#CCF-1527396 and ECCS-1546604.}
\IEEEauthorblockA{Electrical and Computer Engineering Department, George Mason University, Fairfax, VA 22030, USA \\
Email: \{zzhang18, ztian1\}@gmu.edu}
}


%


\maketitle

\begin{abstract}
PhaseLift is a noted convex optimization technique for phase retrieval that can recover a signal exactly from amplitude measurements only, with high probability.
Conventional PhaseLift requires a relatively large number of samples that sometimes can be costly to acquire. 
This paper focuses on some practical applications where the signal of interest is composed of a few Vandermonde components, such as line spectra.
A novel phase retrieval framework, namely ANM-PhaseLift, is developed that exploits the Vandermonde structure to alleviate the sampling requirements. Specifically, the atom set of amplitude-based quadratic measurements is identified, and atomic norm minimization (ANM) is introduced into PhaseLift to considerably reduce the number of measurements that are needed for accurate phase retrieval. The benefit of ANM-PhaseLift is particularly attractive in applications where the Vandermonde structure is presented, such as massive MIMO and radar imaging. 
\end{abstract}

\IEEEpeerreviewmaketitle

\section{Introduction}

Phase retrieval refers to the problem of recovering a signal from only the amplitudes of its linear measurements \cite{candes2015phase}. It appears in a wide range of signal processing applications where sensors cannot measure the phase information, such as X-ray and crystallography imaging, diffraction imaging and microscopy \cite{candes2015phase, waldspurger2015phase}. Phase retrieval can also be applied to applications where the sampled phase information is polluted by unavoidable and inseparable phase errors, such as radar imaging \cite{jaming1999phase}.

In all these problems, complex measurements of a signal $\mathbf{x}\in\mathbb{C}^N$ are sampled from a linear system $\mathbf{Z}\in\mathbb{C}^{M\times N}$, but only the amplitude, or magnitude $|\mathbf{Z x}|^2$, is recorded reliably. Hence, the overall sampling system is nonlinear, and recovering the phase from its amplitude is a non-convex optimization problem. In \cite{candes2015phase, candes2013phaselift, candes2014solving}, a convex relaxation technique named PhaseLift is introduced, which exploits the lifting technique to solve phase retrieval via convex semidefinite programming (SDP). Instead of directly seeking the signal of interest $\mathbf{x}$, PhaseLift solves for a lifted matrix $\mathbf{X}$ that is a proxy for $\mathbf{x}\mathbf{x}^\mathrm{H}$. As such, the quadratic measurements in $\mathbf{x}$ become linear in the semidefinite matrix $\mathbf{X}$, which can be solved via a SDP formulation. Subsequently,  
$\mathbf{x}$ can be easily recovered from $\mathbf{X}$ as its leading eigenvector.

In order to collect adequate information for phase retrieval, PhaseLift usually requires over-sampling, which means that the number of quadratic samples $M$ shall be large enough with respect to the signal size $N$. When some prior information of $\mathbf{x}$ is available, it is possible to reduce the sampling requirements, which is important for those expensive sampling systems. For instance, when $\mathbf{x}$ is sparse, an $\ell_1$-norm regularized sparsity constraint on $\mathbf{x}$ can be incorporated into PhaseLift to reduce the number of samples $M$ needed \cite{li2013sparse, ohlsson2011compressive}. 

This paper considers phase retrieval when the signal of interest $\mathbf{x}$ is known {\em a priori} to be sparsely supported in the frequency domain. Specifically, $\mathbf{x}$ is composed of a few Vandermonde components. This type of signals appears in broad  applications such as DOA estimation and wireless communications. The goal here is to exploit the Vandermonde structure in order to considerably reduce the sampling requirement of PhaseLift. 

An effective way of exploiting the Vandermonde structure for gridless compressed sensing is via atomic norm minimization (ANM), which amounts to a convex SDP formula for line spectrum estimation from a small number of linear measurements \cite{tang2013compressed, candes2014towards}. However, since the ANM approach hinges critically on linearly measurements, it is not directly applicable in phase retrieval where only quadratic measurements are available. To resolve this issue, this paper combines the ANM with the conventional PhaseLift, formulating a non-convex problem describing both $\mathbf{x}$ and $\mathbf{X}$, and then suggests two reformulations to solve it efficiently. Such reformulations allow us to incorporate the ANM into PhaseLift, leading to the proposed ANM-PhaseLift technique. It is shown that the ANM-PhaseLift technique is able to considerably reduce the number of quadratic measurements needed for phase retrieval. 


\section{Background on PhaseLift}
\label{sec:back}


This section reviews the PhaseLift principle for recovering a general-form signal from quadratic measurements \cite{candes2015phase, candes2013phaselift, candes2014solving}. 
Suppose that a signal $\mathbf{x}\in\mathbb{C}^N$ is sampled as quadratic measurements $y_m, m=1, \dots, M$, where
\begin{equation}
	\label{eq:1}
	y_m=|\langle \mathbf{z}_m, \mathbf{x} \rangle|^2, \quad m=1, \dots, M.
\end{equation}
Obviously, $y_m$ only collects amplitude information of $\mathbf{x}$. The recovery of $\mathbf{x}$ from $\mathbf{y}=(y_1, \dots, y_M)^\mathrm{T}$ is a canonical phase retrieval problem.

In (\ref{eq:1}), the quadratic measurements  $\mathbf{y}$ is nonlinear in the unknown $\mathbf{x}$. By introducing a lifted matrix $\mathbf{X} = \mathbf{x}\mathbf{x}^\mathrm{H}$, the PhaseLift technique produces a linear mapping between $\mathbf{y}$ and $\mathbf{X}$ via the following ``lifting'' trick:
\begin{equation}
	\label{eq:1.1}
		y_m =|\langle \mathbf{z}_m, \mathbf{x} \rangle|^2 =\mathbf{z}_m^\mathrm{H} \mathbf{x}\mathbf{x}^\mathrm{H} \mathbf{z}_m =\mathbf{z}_m^\mathrm{H} \mathbf{X} \mathbf{z}_m, \quad \forall m.
\end{equation}
Concisely, the linear mapping in (\ref{eq:1.1}) can be written as $\mathbf{y}=\mathcal{L}(\mathbf{X})$, where $ \mathcal{L}(\cdot)$ is a known linear operator determined by the sampling system $\{\mathbf{z}_m\}_{m=1}^M$. 

Based on the fact that the Hermitian matrix $\mathbf{X}$ is rank-one and semidefinite positive, the PhaseLift technique formulates  the following convex semi-definite programming (SDP) to reconstruct $\mathbf{X}$ and subsequently $\mathbf{x}$ \cite{candes2013phaselift}: 
\begin{equation}
	\label{eq:2}
	\begin{split}
		\min_\mathbf{X} & ~~\mathrm{trace}(\mathbf{X}) \\
		\mathrm{s.t.} & ~~\mathcal{L}(\mathbf{X})=\mathbf{y} \\
		&~~ \mathbf{X}\succeq\mathbf{0}.
	\end{split}
\end{equation}

It is asserted in \cite{candes2015phase, candes2013phaselift, candes2014solving} that, with high probability, $\mathbf{x}\mathbf{x}^\mathrm{H}$ is the unique solution to (\ref{eq:2}) up to a global phase ambiguity, when the number of samples is adequately large with $M\geq C_0 N$ for some constant $C_0$, and the sampling vectors $\{\mathbf{z}_m\}_{m=1}^M$ are  i.i.d. Gaussian random.

\section{Proposed ANM-PhaseLift}
\label{sec:work}

In the absence of linear measurements, PhaseLift in (\ref{eq:2}) generally requires a large number of samples that is sometimes expensive to collect. For some special case of signals, the section presents a new phase retrieval technique that exploits useful \emph{prior information} of $\mathbf{x}$ to alleviate the sampling requirements and improve the noise performance. 


Specifically, we consider a structured signal $\mathbf{x}$ that is composed of a few components in the form of
\begin{equation}
	\label{eq:0}
	\mathbf{x}=\sum_{l=1}^{L} s_l \mathbf{a}(f_l), 
\end{equation}
where $f_l\in[0, 1]$ is the digital frequency of the $l$-th component, and
\begin{displaymath}
	\mathbf{a}(f_l)=\left(1, e^{j2\pi f_l}, e^{j2\pi 2f_l}, \dots, e^{j2\pi (N-1) f_l}\right)^\mathrm{T}
\end{displaymath}  
is the corresponding  manifold vector obeying a Vandermonde structure.
Suppose that $L\ll N$.

The Vandermonde structure has been exploited in gridless compressed sensing via atomic norm minimization (ANM) \cite{tang2013compressed, candes2014towards}. Therein, $\mathbf{x}$ is viewed as a sparse vector over a vector-form atom set 
\begin{equation}
	\mathcal{A}=\{\mathbf{a}(f),\quad \forall f\in[0, 1] \},
\end{equation}
which is of infinite size for the continuously-valued $f$.  
Leveraging the sparsity property of $\mathbf{x}$ over $\mathcal{A}$, we incorporate the ANM into PhaseLift in (\ref{eq:2}) to formulate the following optimization problem: 
\begin{equation}
	\label{eq:4}
	\begin{split}
		\min_{\mathbf{X}, \mathbf{x}} & \left\{\|\mathbf{x}\|_{\mathcal{A}}+\lambda \cdot\mathrm{trace}(\mathbf{X})\right\} \\
		\mathrm{s.t.~} & \mathcal{L}(\mathbf{X})=\mathbf{y} \\
		& \mathbf{X}\succeq\mathbf{0} \\
		& \mathbf{X}=\mathbf{x x}^\mathrm{H},
	\end{split}
\end{equation}
where $\lambda$ is a scalar parameter that controls the sparsity level of  $\mathbf{x}$, balancing the PhaseLift term and ANM term. The formulation in (\ref{eq:4}) for phase retrieval of a line spectrum is termed ANM-PhaseLift. 


\section{Reformulation and Implementation} 

Obviously, the optimization problem in (\ref{eq:4}) is difficult to solve in practice. Main difficulties include:
\begin{itemize}
	\item The Vandermonde structure is implicit in the atomic norm term $\|\mathbf{x}\|_{\mathcal{A}}$, which incurs infinite programming due to the size of $\mathcal{A}$.
	\item While we seek to optimize over $\mathbf{X}$, the original ANM is based on the vector-form atomic norm on $\mathbf{x}$ rather than $\mathbf{X}$. 
	\item The constraint $\mathbf{X}=\mathbf{x x}^\mathrm{H}$ is not convex.
\end{itemize}

The first difficulty can be overcome using the SDP reformulation of atomic norm \cite{tang2013compressed}. 

In detail, 
	if the  wrapped distances of frequencies on the unit circle
		\begin{displaymath}
			\Delta_{\min}=\min_{i\neq j} | f_{i}-f_{j}|, \quad 1\leq i, j\leq L
		\end{displaymath} 
	satisfies the condition
	\begin{equation}
		\label{eq:cs}
		\Delta_{\min}\geq{\frac{1}{\lfloor(N-1)/4\rfloor}},
	\end{equation}
	then it is guaranteed that
	\begin{equation}
		\begin{split}
			\|\mathbf{x}\|_\mathcal{A}=\min_{\mathbf{u}, v}~ & \left\{v+\frac{1}{N}\mathrm{trace}\big(\mathbf{T}(\mathbf{u})\big)\right\} \\
			& \mathrm{s.t.~} ~\left(\begin{array}{cc}
			v & \mathbf{x}^\mathrm{H} \\
			\mathbf{x} & \mathbf{T}(\mathbf{u})
			\end{array}\right)\succeq \mathbf{0}.
	\end{split}
	\end{equation}
	where $\mathbf{T}(\mathbf{u})$ is a Hermitian Toeplitz matrix constructed from its first row $\mathbf{u}$.

Hence, if $\{f_l\}_{l=1}^L$ are well separated as indicated in (\ref{eq:cs}),
it is straightforward to rewrite (\ref{eq:4}) as the following formulation
\begin{eqnarray}
	\label{eq:6}
	\min_{\mathbf{u}, v, \mathbf{X}, \mathbf{x}}~ & \left\{v+\frac{1}{N}\mathrm{trace}\big(\mathbf{T}(\mathbf{u})\big)+\lambda\cdot \mathrm{trace}(\mathbf{X})\right\} \\
		\nonumber\mathrm{s.t.~} & ~~\mathcal{L}(\mathbf{X})=\mathbf{y} \\
		\nonumber& ~~\mathbf{X}\succeq\mathbf{0} \\
		& ~~ \left(\begin{array}{cc}
		v & \mathbf{x}^\mathrm{H} \\
		\mathbf{x} & \mathbf{T}(\mathbf{u})
		\end{array}\right)\succeq \mathbf{0} \label{eq:61} \\
		& ~~ \mathbf{X}=\mathbf{x x}^\mathrm{H} \label{eq:62}.
\end{eqnarray}

Equation (\ref{eq:6}) is the basic expression of ANM-PhaseLift. In the following subsections, we will focus on the non-convex constraint (\ref{eq:62}) to reformulate (\ref{eq:6}) to a computationally feasible way.

\subsection{BMI Reformulation}

It can be observed that among all constraints, only (\ref{eq:61}) is not involved with $\mathbf{X}$. 
It gives rise to the aforementioned second difficulty, in which (\ref{eq:61}) results from the ANM term for linear measurements, but we prefer to formulate the problem in terms of $\mathbf{X}$ for quadratic measurements. This suggests us to rewrite (\ref{eq:61}) in order to cancel out (\ref{eq:62}). 


According to Schur complement condition for positive semidefinite matrices, we have the following equivalence of (\ref{eq:61}):
\begin{equation}
	\begin{split}
		& \left(\begin{array}{cc}
		v & \mathbf{x}^\mathrm{H} \\
		\mathbf{x} & \mathbf{T}(\mathbf{u})
		\end{array}\right)\succeq \mathbf{0} \\
		\Leftrightarrow & ~ v>0, \quad\mathbf{T}(\mathbf{u})-\mathbf{x} v^{-1} \mathbf{x}^\mathrm{H} \succeq \mathbf{0} \\
		\Leftrightarrow & ~ v>0, \quad v\cdot \mathbf{T}(\mathbf{u})-\mathbf{X}\succeq \mathbf{0}.
	\end{split}
\end{equation}

Hence, (\ref{eq:6}) can be rewritten as
\begin{equation}
\label{eq:63}
	\begin{split}
		\min_{\mathbf{u}, v, \mathbf{X}}~ & \left\{v+\frac{1}{N}\mathrm{trace}\big(\mathbf{T}(\mathbf{u})\big)+\lambda\cdot \mathrm{trace}(\mathbf{X})\right\} \\
		\mathrm{s.t.~} & ~~\mathcal{L}(\mathbf{X})=\mathbf{y} \\
		& ~~\mathbf{X}\succeq\mathbf{0} \\
		& ~~ v>0 \\
		& ~~ v\cdot \mathbf{T}(\mathbf{u})-\mathbf{X}\succeq \mathbf{0}.
	\end{split}
\end{equation}
Evidently, (\ref{eq:63}) is no longer involved with $\mathbf{x}$, and hence the constraint in (\ref{eq:62}) can be removed.

The reformulation in (\ref{eq:63}) contains a bilinear matrix inequality (BMI) constraint $v\cdot \mathbf{T}(\mathbf{u})-\mathbf{X}\succeq \mathbf{0}$. Usually bilinear constraint is difficult to deal with. An intuitive solution is to perform two-step iterations, that is, (\ref{eq:63}) is optimized over $v$ and $\mathbf{T}(\mathbf{u})$ in an alternating manner, as listed in Algorithm \ref{alg:1}. 

\begin{algorithm}
	\caption{Iterative BMI ANM-PhaseLift}
	\label{alg:1}
	\begin{algorithmic}[1]
		\STATE Set the initial values $\mathbf{u}^0$, $v^0$ and $\mathbf{X}^0$;
		\STATE $i\leftarrow 0$;
		\REPEAT
			\STATE Fix $\mathbf{u}^i$, optimize (\ref{eq:63}) over $v$ and $\mathbf{X}$ (\emph{convex});
			\STATE Fix $v^{i+1}$, optimize (\ref{eq:63}) over $\mathbf{u}$ and $\mathbf{X}$ (\emph{convex});
			\STATE $i\leftarrow i+1$;
		\UNTIL{convergence}
		\RETURN $\hat{\mathbf{u}}=\mathbf{u}^i$ and $\hat{\mathbf{X}}=\mathbf{X}^i$;
		\STATE Find $\hat{\mathbf{x}}$ as the leading eigenvector of $\hat{\mathbf{X}}$, or find $\hat{\mathbf{f}}$ directly from $\hat{\mathbf{u}}$ via Vandermonde decomposition, if desired.
	\end{algorithmic}
\end{algorithm}

\subsection{Convex Reformulation}

Another way of reformulating (\ref{eq:6}) is to convexify (\ref{eq:62}) directly. Specifically, the constraint in (\ref{eq:62}) is equivalent to the following two constraints \cite{doi:10.1080/002071700219803}:
\begin{equation}
	\label{eq:71}
	\mathbf{X}\succeq\mathbf{x}\mathbf{x}^\mathrm{H},
\end{equation}   
\begin{equation}
	\label{eq:72}
	\mathrm{trace}(\mathbf{X})-\sum_{n=1}^{N} |x_n|^2 \leq 0.
\end{equation}

Since the constraint in (\ref{eq:72}) is concave, \cite{boyd1997semidefinite} suggests that we can further drop this constraint as a relaxation, and (\ref{eq:71}) can be equivalently written as a SDP constraint
\begin{equation}
	\label{eq:73}
	\left(\begin{array}{cc}
		1 & \mathbf{x}^\mathrm{H} \\
		\mathbf{x} & \mathbf{X}
	\end{array}\right)\succeq\mathbf{0}.
\end{equation}

Directly replacing the constraint in (\ref{eq:62}) by (\ref{eq:73}) will make the relaxation too loose. Alternatively, we combine (\ref{eq:73}) with another SDP constraint (\ref{eq:61}), and hence (\ref{eq:6}) is relaxed to the following convex optimization after renaming $\mathbf{x}$ by $\mathbf{w}$: 
\begin{equation}
	\label{eq:74}
	\begin{split}
	\min_{\mathbf{u}, v, \mathbf{X}, \mathbf{w}}~ & \left\{v+\frac{1}{N}\mathrm{trace}\big(\mathbf{T}(\mathbf{u})\big)+\lambda\cdot \mathrm{trace}(\mathbf{X})\right\} \\
	\mathrm{s.t.~} & ~~\mathcal{L}(\mathbf{X})=\mathbf{y} \\
		& ~~\mathbf{X}\succeq\mathbf{0} \\
	& ~~\left(\begin{array}{cc}
	 v & \mathbf{w}^\mathrm{H} \\
	\mathbf{w} & \mathbf{T}(\mathbf{u})+\mathbf{X}
	\end{array}\right)\succeq \mathbf{0}.
	\end{split}
\end{equation}
The reformulation in (\ref{eq:74}) can be solved via a popular convex optimization toolbox. The phase retrieval result $\hat{\mathbf{x}}$ or the line spectrum estimate $\hat{\mathbf{f}}$ can be further retrieved from the estimated $\hat{\mathbf{X}}$ or $\mathbf{T}(\hat{\mathbf{u}})$, similar to Step 9 in Algorithm 1.

\subsection{Remarks}

\noindent \textbf{\emph{Remark 1:}} The convex reformulation in (\ref{eq:74}) is a relaxation of (\ref{eq:6}) and is subject to a performance degradation compared with the BMI reformulation (\ref{eq:63}). But (\ref{eq:74}) is much more computationally efficient because it obviates the bilinear constraint, and the performance loss is usually very small. We will show this in the next section via simulations.

\noindent \textbf{\emph{Remark 2:}} In line spectrum estimation applications, oftentimes the goal is to estimate the frequencies $\{f_l\}_{l=1}^L$, not $\mathbf{x}$ itself. In this case, we can directly obtain the frequency information from $ \mathbf{T}(\mathbf{u})$ or $\mathbf{u}$ via Vandermonde decomposition, without having to recover $\mathbf{x}$ from $\mathbf{X}$. This is evident from the structure of $ \mathbf{T}(\mathbf{u})$:
\begin{equation} 
\begin{split}
\mathbf{T}(\mathbf{u})=&\mathbf{A}(\mathbf{f}) \mathbf{D} \mathbf{A}^{\mathrm{H}}(\mathbf{f}), \\
& \mathbf{D} \succeq \mathbf{0} \mathrm{~is~diagonal}. 
\end{split}
\end{equation}
Note that the frequency information is not affected by the global phase ambiguity of PhaseLift. That is, we can not only find the solution to $\mathbf{x}$ (phase retrieval problem), but also reveal the frequency components $\{f_l\}_{l=1}^L$ directly (line spectrum estimation problem).

\noindent \textbf{\emph{Remark 3:}} In the presence of noise, it is straightforward to reformulate (\ref{eq:6}) with Least Squares regularization on the noisy measurements $\mathbf{y}$. We take the BMI reformulation as an example:
\begin{equation}
\label{eq:7}
\begin{split}
\min_{\mathbf{u}, v, \mathbf{X}} & \left\{v+\frac{1}{N}\mathrm{trace}\big(\mathbf{T}(\mathbf{u})\big)+\lambda\cdot \mathrm{trace}(\mathbf{X})\right.\\
&~~ + \delta\cdot \|\mathcal{L}(\mathbf{X})-\mathbf{y} \|_2^2 \Bigg\} \\
\mathrm{s.t.} 
&~~ \mathbf{X}\succeq\mathbf{0} \\
&~~ v>0 \\
&~~ v\cdot \mathbf{T}(\mathbf{u})-\mathbf{X}\succeq \mathbf{0},
\end{split}
\end{equation}
where the scalar $\delta$ is a regularization parameter.

\section{Simulations}
\label{sec:sim}

Monte Carlo simulations are conducted to illustrate the advantages of ANM-PhaseLift over the conventional PhaseLift for phase retrieval and line spectrum estimation, when the Vandermonde structure is present. Throughout, $L=2$, $N=8$, and the number of samples $M$ varies from 4 to 32. The sampling vectors $\{\mathbf{z}_m\}_{m=1}^M$ are i.i.d. normal Gaussian distributed, and $\{s_l\}_{l=1}^L, \{f_l\}_{l=1}^L$ are arbitrarily selected. Both the BMI reformulation and convex reformulation are tested, with the comparison to conventional PhaseLift for benchmark.

\subsection{Phase Retrieval}

\begin{figure}
	\centering
	\includegraphics[width=\columnwidth]{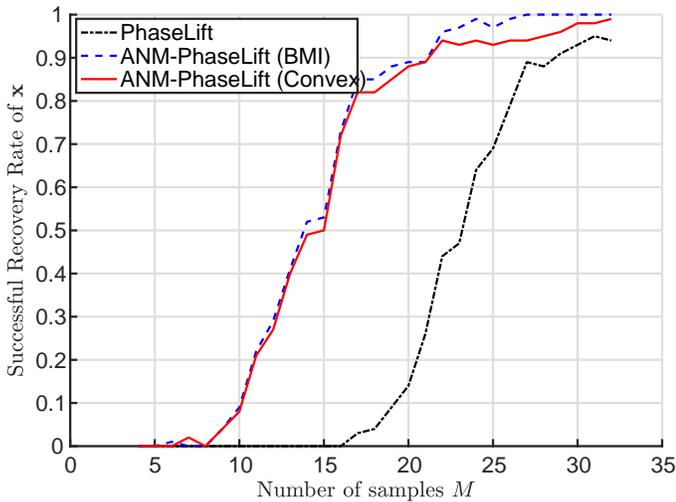}\vspace{-8pt}
	\caption{The success rate of phase retrieval.}
	\label{fig:1}
\end{figure}

Figure \ref{fig:1} tests the phase retrieval performance, which depicts the success rate of reconstructing the signal $\mathbf{x}$ from quadratic measurements. It corroborates that the proposed ANM-PhaseLift technique reduces the number of samples $M$ needed for successful signal recovery, and attains better error performance given $M$, especially for the BMI reformulation. 

In the phase retrieval test, the convex reformulation of ANM-PhaseLift shows a slightly performance degradation compared with the BMI reformulation due to its relaxation. Obviously, both reformulations of ANM-PhaseLift performs much better than the conventional PhaseLift.

\subsection{Line Spectrum Estimation}

\begin{figure}
	\centering
	\includegraphics[width=\columnwidth]{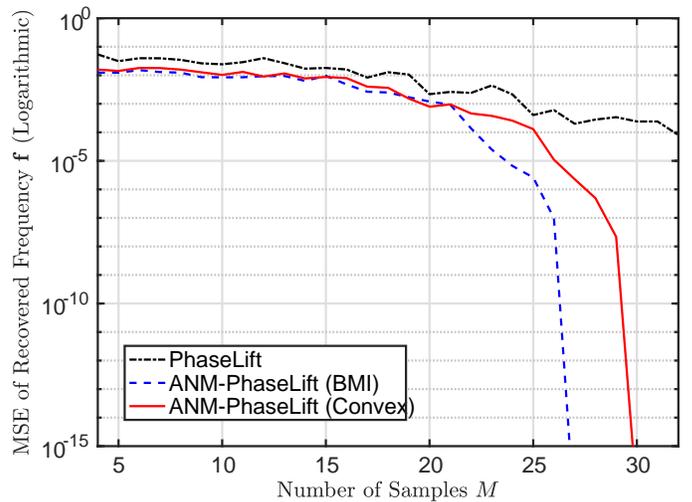}\vspace{-8pt}
	\caption{MSE performance of frequency estimation (Logarithmic scale).}
	\label{fig:2}
\end{figure}

Figure \ref{fig:2} tests the performance of line spectrum estimation by comparing the mean square error (MSE) of the recovered frequencies $\{\hat{f}_l\}_{l=1}^L$ for the following two schemes: \vspace{-0.03in}
\begin{itemize}
\itemsep -0.01in
	\item In ANM-PhaseLift, $\{\hat{f}_l\}_{l=1}^L$ are directly obtained from $\mathbf{T}(\hat{\mathbf{u}})$ via Vandemonde decomposition, without having to estimate $\hat{\mathbf{x}}$.
	\item  In conventional PhaseLift, $\hat{\mathbf{X}}$ is obtained from (\ref{eq:4}) first, and then the ANM is applied to estimate the frequencies as described in \cite{tang2013compressed}.  \vspace{-0.03in}
\end{itemize}

It is shown that both reformulations of ANM-PhaseLift exhibit evident performance advantages for frequency estimation compared with conventional PhaseLift, and the performance gap between two reformulations is also very small. 
Further, ANM-PhaseLift is computationally more efficient, because it avoids the step of retrieving the phase of $\hat{\mathbf{x}}$.

\section{Summary}
\label{sec:con}

This work presents a novel convex optimization framework for phase retrieval, when the signal of interest possesses a Vandermonde structure.  By properly combining both PhaseLift and ANM, the proposed ANM-PhaseLift is able to considerably reduce the number of quadratic measurements needed for accurate phase retrieval. It can also efficiently retrieve the frequency information of the signal components, without having to recovering the signal itself. For future work, it is important to delineate the theoretical bound on the required number of samples for a given sparsity of $\mathbf{x}$ over $\mathcal{A}$. 

\bibliographystyle{IEEEbib}
\bibliography{overview}

\end{document}